\documentclass{PoS}
\usepackage[authoryear,square]{natbib}
\bibpunct{(}{)}{;}{a}{}{,}

\usepackage{graphicx,times}
\usepackage{epsf}
\usepackage{amsmath}
\usepackage[title]{appendix}

\newcommand{\skipthis}[1]{}

\def\deg{{\rm deg}}

\def\Mpc{{\rm Mpc}}
\def\Gpc{{\rm Gpc}}

\def\MHz{{\rm MHz}}

\title{Topology of neutral hydrogen distribution 
with the Square Kilometer Array}

\ShortTitle{HI topology}

\author{Yougang Wang$^1$, Yidong Xu$^1$, Fengquan Wu$^1$, \speaker{Xuelei Chen}$^1$,
Xin Wang$^2$,
Juhan Kim$^3$,
Changbom Park$^3$,
Khee-Gan Lee$^4$,
Renyue Cen$^4$
\\

$^1$National Astronomical Observatory, Chinese Academy of Sciences, Beijing, China;\\
$^2$Department of Physics and Astronomy, The Johns Hopkins University, Baltimore, USA;\\
$^3$School of Physics, Korean Institute for Adavanced Studies, Seoul, Korea;\\
$^4$Department of Astrophysical Sciences, Princeton University, Princeton, USA\\
E-mail: \email{wangyg@bao.ac.cn, xuelei@cosmology.bao.ac.cn}
}

\abstract{
Morphology of the complex HI gas distribution can be quantified by statistics
like the Minkowski functionals, and can provide a way to statistically study 
the large scale structure in the HI maps both at low redshifts, and 
during the epoch of reionization (EoR). 
At low redshifts, the 21cm emission traces the underlying matter distribution. 
Topology of the HI gas distribution, as measured by the genus, 
could be used as a ``standard ruler''. This enables the 
determination of distance-redshift relation 
and also the discrimination of various models of dark energy and of modified 
gravity. The topological analysis is also sensitive to certain primordial 
non-Gaussian features. Compared with two-point statistics, the topological 
statistics are more robust against the nonlinear gravitational evolution, 
bias, and redshift-space distortion. The HI intensity map observation naturally
 avoids the sparse sampling distortion, which is an important systematic in 
optical galaxy survey. The large cosmic volume accessible to SKA would provide 
unprecedented accuracy using such a measurement.
During the EoR, topology can be a powerful and intuitive tool to 
distinguish among the different evolutionary 
stages of reionization, where the ionized regions make up a significant 
fraction of the volume. Furthermore, it can also discriminate among various 
reionization models. The genus curves evolve 
during cosmological reionization, and for different reionization scenarios, 
the topology of the HI gas distribution can be significantly different even 
if the global ionization fractions are the same. It can provide clear and 
intuitive diagnostics for how the reionization takes place, and indirectly 
probes the properties of radiation-sources.  
In this brief chapter we will describe the scientific background of the 
topology study, and forecast the potential of the SKA for measuring cosmological
parameters and constraining structure formation mechanism through the study of
topology of HI gas distribution.

}

\FullConference{
Advancing Astrophysics with the Square Kilometre Array\\
June 8-13, 2014\\
Giardini Naxos, Italy}

\begin{document}

\section{Topology study of HI distribution }
Topology has been introduced in cosmology to describe the properties 
of the large-scale structure (LSS), and to
test the non-Gaussianity of the primordial density fluctuations
\citep{1986ApJ...306..341G,1986ApJ...309....1H,2013JKAS...46..125P}. 
Application of LSS topology study has been extended to
measurement of cosmological parameters and constraining galaxy formation
mechanism \citep{2005ApJ...633....1P, 2010ApJS..190..181C, 2013ApJS..209...19C}. The LSS
topology is also a cosmological invariant that can be used to reconstruct
the expansion history of the universe 
\citep{2010ApJ...715L.185P,2011MNRAS.412.1401Z, 2012ApJ...747...48W, 2013arXiv1310.4278S,2014MNRAS.437.2488B}.
Mathematically, geometry of the excursion regions with density above a
threshold can be characterized by the Minkowski Functionals (MFs) 
\footnote{In three dimensions, the four Minkowski functionals for the isodensity contours
are  the volume fraction $V_0(\nu)$, surface area $V_1(\nu)$, mean curvature $V_2(\nu)$ and 
Euler characteristics $V_3(\nu)=\chi(\nu)$, which is related to the genus by $\chi=2-2g$.}
\citep{1994A&A...288..697M, 2012MNRAS.423.3209P, 2006ApJ...653...11H,2008MNRAS.385.1613H}. 
The genus, related to one of the Minkowski functionals, is given by the number 
of holes minus the number of isolated regions in the iso-density contour 
surfaces of the smooth density field, and is a measure of topology that 
quantifies the connectivity of the contour surfaces \citep{2013JKAS...46..125P}. 
All the MFs are analytically known for Gaussian fields, the
genus per unit volume as a function of the threshold density 
$\nu=\delta/\sigma$ is given by 
\begin{equation}
G(\nu) =  A (1-\nu^2) e^{-\nu^2/2}, \qquad
{ A} = 
\frac{1}{4 \pi^2}\left(\frac{\langle k^2 \rangle}{3} \right)^{3/2}
\label{eqn:genus_A}
\end{equation}
 The topology of the isodensity contours is insensitive to the systematic
effects such as redshift-space distortion, non-linear evolution, sparse
sampling \citep{2005ApJ...633....1P,2010ApJ...715L.185P}. Therefore, any deviation from the Gaussian prediction 
is evidence for non-Gaussianity, and can constrain the mechanism for primordial 
non-Gaussianity. The deviation can be caused not just by the three-point function but 
by all the high-order moments of the density field. Therefore, the study of 
non-Gaussianity using MFs is complementary to the approach of using three-point correlation 
function or bispectum, and is in principle able to detect general forms of non-Gaussianity.
The Minkowski functionals measured from the HI 21cm map can characterize 
the complex distribution of the neutral hydrogen, both in the large-scale 
structure at low redshifts, and during the epoch of reionization (EoR) at high redshifts.

\subsection{Primordial non-Gaussianity}
The primordial perturbation may deviate from the Gaussian random field 
during the cosmological phase transition triggered by the spontaneous 
breakdown of symmetry
creating topological defects such as the cosmic monopole, string, wall, and textures. 
Various non-standard inflationary scenarios which violate the single-field 
and slow-rolling condition may also produce significant primordial non-Gaussianity.
It is usual to parameterize the non-Gaussian features generated by the inflation model as
$\Phi(x) = \phi_G(x) + f_{\rm NL}\left[ \phi^2_G(x) - \left<\phi^2_G\right>\right]$
where $\phi_G$ is the Gaussian distribution of potential. For this form of 
primordial non-Gaussianity, the genus curve deviates from the 
Gaussian model, and up to second order perturbations it is given 
by \cite{ 2006ApJ...653...11H,2008MNRAS.385.1613H}.
\begin{equation}
\Delta(\nu)\equiv \delta\left(\frac{G}{A}\right)
=-e^{-\nu^2/2}\times f_{\rm NL}\left[(S_{\rm pri}^{(1)}-S_{\rm pri}^{(0)})H_3(\nu_f)+
(S_{\rm pri}^{\prime(2)}-S_{\rm pri}^{\prime(0)})H_1(\nu)\right]\sigma_{0}
\end{equation}
where $S_{\rm pri}^{(a)}$ is the skewness parameter \citep{2003ApJ...584....1M}.
The measured genus curve can then be used to constrain $f_{NL}$. However, even if the 
non-Gaussianity can not be parameterized in this form, it would still affect the 
genus curve, so the topology can be used to probe more general forms
of primordial non-Gaussianity.

\subsection{Dark Energy and Modified Gravity}

\begin{figure}[!htbp]
\begin{center}
\includegraphics[width=0.59\textwidth,height=0.4\textwidth]{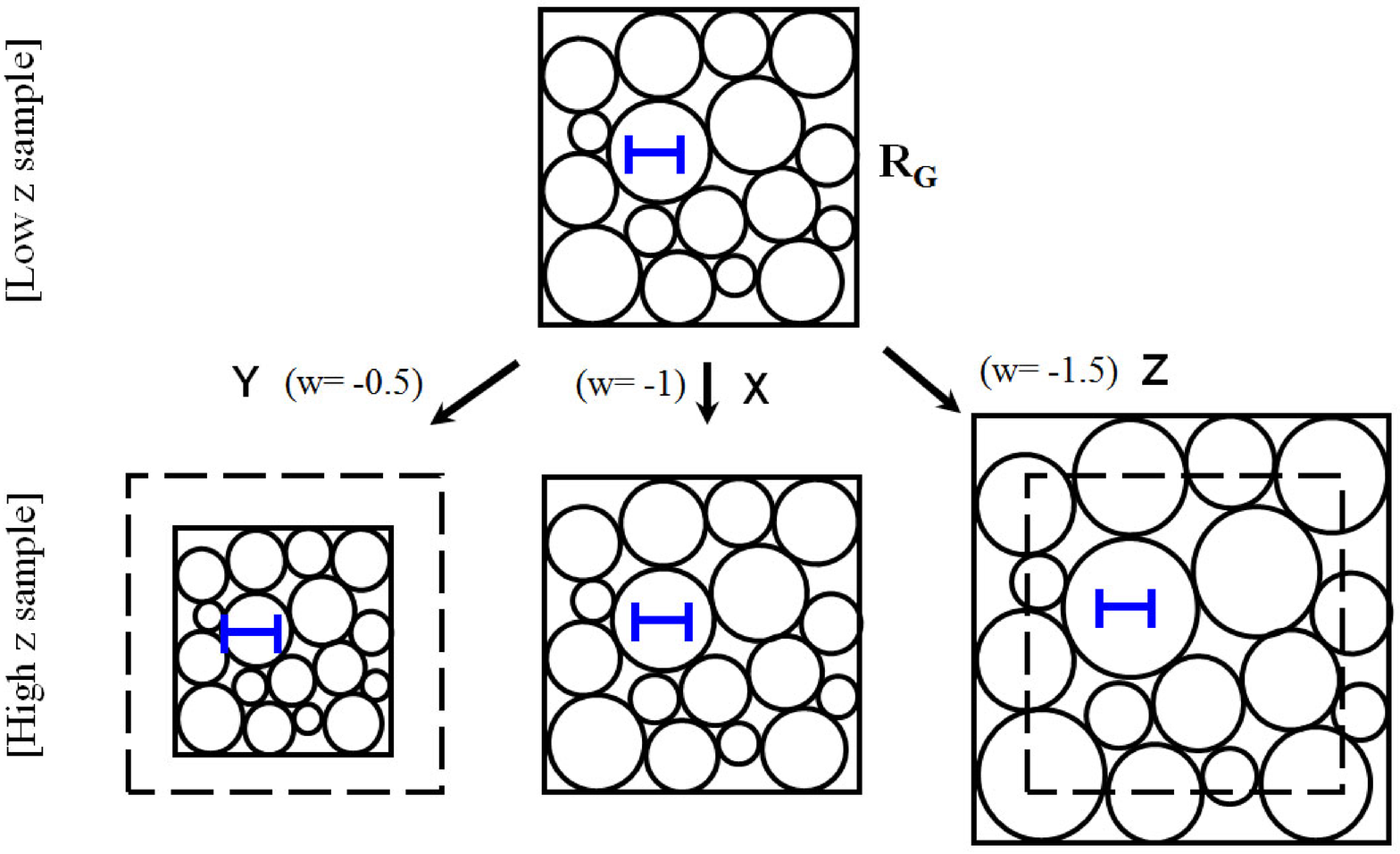}
\includegraphics[width=0.4\textwidth]{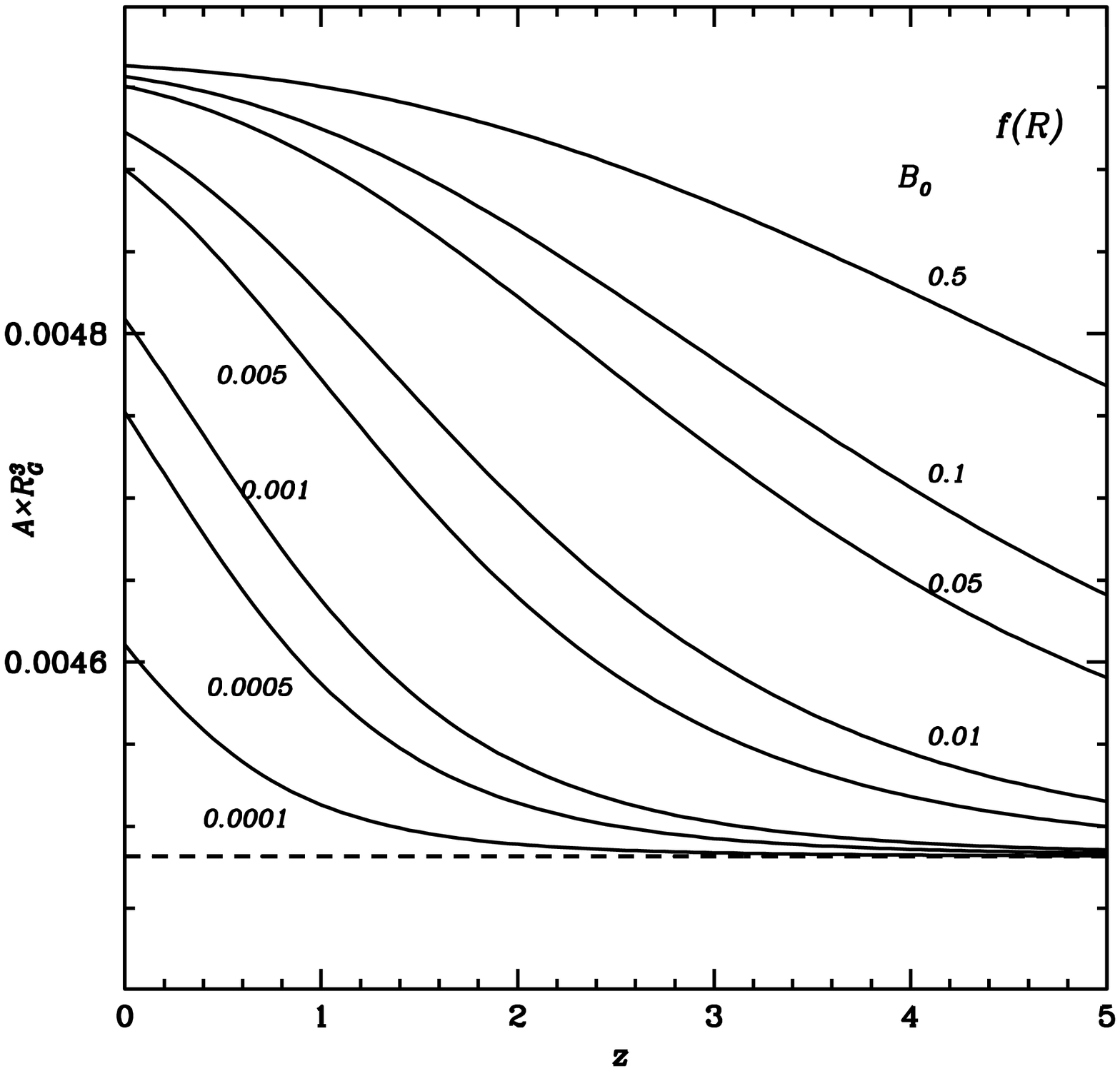}
\end{center}
\caption{\label{fig:stdruler} Left: The number of structures as cosmic standard ruler.
Right: the redshift evolution of genus amplitude
for $f(R)$ and phenomenological modified gravity models with different $B_0$ values. }
\end{figure}

The LSS topology is relatively stable with the growthing of structure.
Eq.~(\ref{eqn:genus_A}) shows that ${A}$  measures the 
slope of the power spectrum around the smoothing scale $R_G$,
and for linear growth  $A$ would be conserved \citep{2005ApJ...633....1P}, so
as long as the same smoothing scale is used, the 
same comoving volume at different redshifts would enclose nearly
the same amount of structures, as illustrated in the left panel of Fig.\ref{fig:stdruler} 
This fact can be used to measure the redshift-distance relation $r(z)$ \citep{2010ApJ...715L.185P}.
Since the topology of the structure is not scale-free, 
the genus enclosed in a wrongly sized volume and smoothed with a wrong scale would
lead to deviation from the actual one, 
\begin{eqnarray}
 {A} _Y(z, R_{{G},Y}) R_{{G},Y}^3 =  
{A} _X(R_{{G},X}) R_{{G},X}^3,
\end{eqnarray}
where $R$ is the volume, $Y$ is the adopted cosmology parameters while $X$ is the true 
cosmology. The smoothing scale $R_{{G}}$ for different cosmologies 
are related by 
$(R_{{G},X}/ R_{{G},Y} )^3= (D_A^2/H)_X/(D_A^2/H)_Y ,$
where $D_A, H$ are the angular diameter distance and Hubble expansion rate respectively.
Utilizing this effect, topology of the large scale structure 
can serve as a standard ruler in cosmology. This technique has been applied to 
constrain dark energy model \citep{2014MNRAS.437.2488B, 2011MNRAS.412.1401Z}.

For modified gravity models, the growth rate is different from the 
general relativity, and in some cases
it is scale-dependent. This will induce changes in the genus curve. 
In the right panel of Fig.\ref{fig:stdruler} the redshift evolution 
of the genus amplitude for the  
$f(R)$ theory and some phenomenological modified gravity models 
are shown.  Here the modified gravity model parameter  $B_0$ is the present day value of the function $B(a)$,  which is the square of
the Compton scale given by $B(a)\equiv \frac{f_{RR}}{1+f_{R}}{R}^\prime\frac{H}{H^\prime} $, where  $ f_{R} $ and $f_{RR}$ are the first and second
derivatives of $f(R)$, with $R$ being the Ricci scalar.

So the topological 
measurements can also be used to constrain the modified gravity 
models \citep{2012ApJ...747...48W}. Compared with direct measurement
using the growth factor, the topological measurement may be
 more robust, as they suffer less from the 
effect of bias, redshift distortion, and non-linearity.

\subsection{Reionization}
\label{intro:reion}
The topology analysis provides a direct and sensitive 
probe for the detailed process of cosmic reionization \citep{2008ApJ...675....8L, 2014JKAS...47...49H}. 
Using the genus of HI density contours as a quantitative measure of topology, 
the reionization process of the intergalactic medium (IGM) can be divided into four distinct 
topological phases for the standard scenario of reionization: 
(1) pre-reionization, before the formation of any HII bubbles, the HI distribution 
reflects the primodial density fluctuations, the genus 
curve is consistent with a Gaussian density distribution, 
and topology remains unchanged as the HI density evolves linearly; 
(2) pre-overlap, characterized by a topology 
dominated by isolated HII bubbles.  At the earlier stages of this phase, 
the number of HII bubbles increases gradually with time, 
resulting in an increase in the amplitude of the genus curve, while 
at the later stages when HII bubbles start to merge, the genus curve goes through a turnover 
and subsequently decreases in amplitude. (3) overlap, 
during which the amplitude of the genus curve drops significantly as 
the HII bubbles rapidly merge; 
(4) postoverlap, when the IGM is almost ionized, 
and the evolution of the genus curve is consistent with an decreasing number 
of isolated neutral islands.
The measurement of the genus curve can be used to characterize the evolutionary 
stages of the neutral topology and to identify the redshifts 
at which the different stages of reionization occur, thus distinguishing 
different reionization scenarios. This in turn provides information on 
the nature and properties of the early luminous sources.

\section{The SKA performance}
The observation of HI emission is presently limited to $z<0.2$ by the sensitivity of 
existing telescopes. The SKA, with its great increase in sensitivity, 
will allow us to observe the HI at
much higher redshifts, and probe the topology of HI distribution 
with unprecedented capability. The SKA will be built in two 
stages (Dewdney et al. 2013; Braun), the SKA-1 will consist three arrays: SKA-mid,
SKA-sur and SKA-low, the features are listed in Table.\ref{tab:SKA_array};
The SKA2 is to be designed in the future, but should be about 10 times larger in 
mid-frequency and 4 times larger in the low frequencies.

\begin{table}[ht]
\begin{center}
\caption{\label{tab:SKA_array} The three SKA1 arrays.}
\begin{tabular}{|c|c|c|}
\hline
\hline
Array & configuration & relevant band \\
\hline
SKA-mid & $190 \times 15$m dish + 64 MeerKAT 13.5m dish &  band 1(0.35-1.05GHz), band 2(0.95-1.76GHz)\\ 
\hline
SKA-sur & $60 \times 15$m dish + 36 ASKAP 12m dish (PAF) & band 1 (0.35-0.5) and 
band 2 0.65-1.67 GHz. \\
\hline  
SKA-low & 250,000 log periodic antenna in 866 stations & 50-350 MHz\\
\hline
\hline
\end{tabular}
\end{center}
\end{table}

The SKA-low is designed for observation of the high redshifts of the EoR, 
while SKA-mid for the low-to-mid-redshifts of post-reionization Universe, 
and SKA-sur with its larger FoV is suitable for survey of large area. 
All of these can be used for HI topology study. The
topological analysis can be applied on data from all HI surveys. 
We expect that large HI surveys will be one of the key science projects of 
SKA, being carried out with multiple applications, and the HI topology 
can be studied in conjunction with the other projects.

A classic approach is the HI galaxy survey where the 
individual galaxies are observed as objects whose HI content exceeds detection threshold: 
\begin{equation}
S_{lim}=N_{th} \frac{k T_{sys}}{A_{eff} \sqrt{\Delta\nu t}}
=N_{th} \frac{\rm SEFD}{\sqrt{\Delta\nu t}}
\end{equation}
where $N_{th}$ is a preset threshold value, e.g. 5 or 10. The SKA is potentially 
capable of surveying a billion of galaxies, with SKA-1 
it is capable of surveying $10^{7-8}$ galaxies\citep{2005MNRAS.360...27A,2009astro2010S.219M} at 
$z$ up to $\sim1.5$. But even smaller surveys, e.g. those conducted with SKA-1 early science, or 
for small deep field, would for the first time reveal the HI distribution 
at redshifts much beyond our current limit, and the topological analysis 
will help us to understand quantitatively the characteristics of HI 
distribution and redshift evolution, 
as has been down for low redshift optical galaxies \citep{2010ApJS..190..181C,2013ApJS..209...19C}.

Another approach is intensity mapping (IM): the sky is mapped at 
low angular resolution, such that individual galaxies can not 
be distinguished, but data of HI large scale distribution 
can still be obtained \citep{2008PhRvL.100i1303C}.
For the EoR experiments, including that of the SKA-low, due to the
low angular resolution, this will also be the mode of observation.
A few dedicated experiments, such as the
CHIME\footnote{http://chime.phas.ubc.ca/} and Tianlai \footnote{http://tianlai.bao.ac.cn/}
are designed to conduct IM observations.
While the SKA-mid or SKA-sur are not designed for this,
it has been proposed that IM surveys may be 
conducted with the dishes used individually with output autocorrelation, 
while being calibrated using the interferometry  (Santos 2014). Whether this would 
work still needs to be tested in the field, but if successful, 
it would allow higher sensitivity on interested scales
at the price of lower angular resolution.
\begin{table}[ht]
\begin{center}
\caption{\label{tab:SKA_parameters} SKA survey parameters utilized in this paper.}
\begin{tabular}{|c|c|c|c|}
 \hline
galaxies surveys & survey area ($\deg^2$) &  integration time ($h$) & flux limit (${\rm mJy}$)\\
\hline
SKA1-Mid & $1,000$ ($5,000$) & $5,000$ ($25,000$)  & $0.13$\\
SKA2-Mid & $1,000$ ($5,000$) & $5,000$ ($25,000$)  & $0.05$ \\
SKA1-Sur & $30,000$ & $7,500$ ($50,000$) & $0.4$ ($0.15$) \\
SKA2-Sur & $30,000$ & $7,500$ & $0.05$ \\
\hline
\hline
HI intensity mapping & survey area ($\deg^2$) &  integration time ($h$) & frequency band ($\MHz$) \\
\hline
SKA1-Mid-SD  & $30,000$ & $10,000$  & $350-1050$ \\
SKA1-Mid-Int & $30,000$ & $10,000$  & $350-800$ \\
SKA2-Mid-Int & $30,000$ & $10,000$  & $350-950$ \\
\hline
\end{tabular}
\end{center}
\end{table}

In Table\ref{tab:SKA_parameters} we list both the HI galaxy and IM
survey parameters we adopted for our forecast on dark energy and modified gravity. 
For SKA-low, as the design is still
very uncertain, we do not make a full forecast here, but only discuss the potential 
of the measurement.

\section{Large Scale Structures }
After EoR, with the IGM highly ionized, the neutral hydrogen resides 
mainly in galaxies, and the HI topology traces out the LSS. We can use this
observation to constrain the primordial non-Gaussianity, 
 dark energy, and modified gravity.

\begin{figure}
\centering
\includegraphics[width=0.65\textwidth,height=0.5\textwidth]{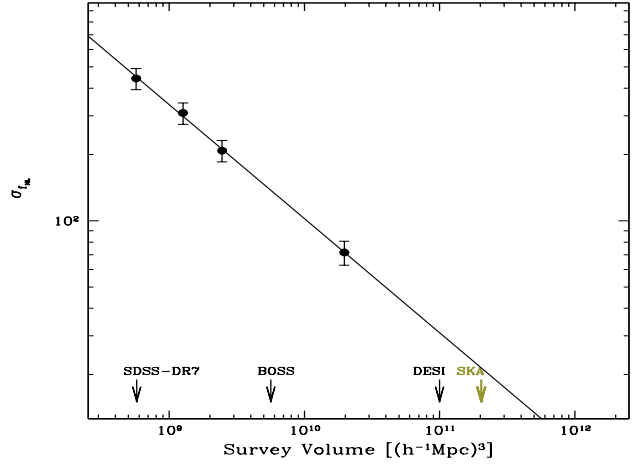}
\caption{
The measurement error of $f_{\rm NL}$ from genus curves.
We mark the survey volumes of previous (SDSS-DR7, BOSS)
and forth-coming (DESI \& SKA) surveys.
}
\label{fnl}
\end{figure}

For primordial non-Gaussianity, Fig.\ref{fnl} shows the $f_{\rm NL}$ constraints from 
measurement of genus curve in the Horizon Run III simulation, 
which has a volume of $10.8^3 h^{-3}\Gpc^3$.
We employed a Gaussian smoothing of $R_G=22 h^{-1}\Mpc$ to build
a density field from subcubes, and compare the result for four survey volumes: SDSS, 
BOSS, DES and SKA. If the SKA-2 will observe galaxy distribution
out to $z=3$ with a solid angle of 20,000 deg$^2$, then using the 
genus-curve we can measure $f_{\rm NL}$ with an error, $\sigma(f_{\rm NL})=20$. 
We should note that this is comparable to the CMB limits, and the genus measurement can 
detect more general forms of non-Gaussianity.

Using the genus curve as standard ruler, dark energy equation of state parameters
$w_0, w_a$ can be constrained. As shown in Table (\ref{tab:SKA_parameters}), we 
consider a relatively deep survey using SKA-Mid compared with 
shallow but wider surveys for SKA-Sur.

\begin{figure}[!htbp]
\begin{center}
\includegraphics[width=0.49\textwidth]{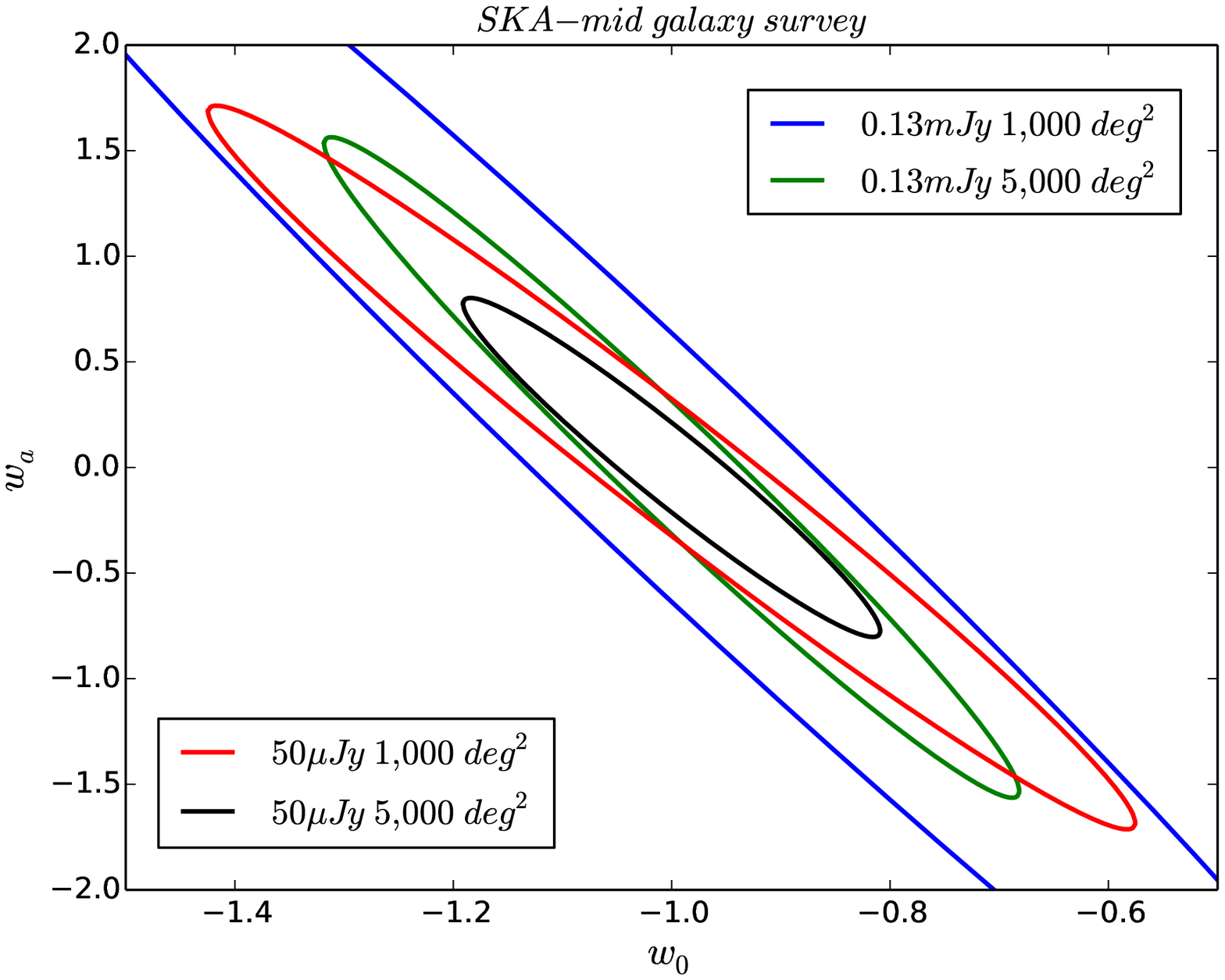}
\includegraphics[width=0.49\textwidth]{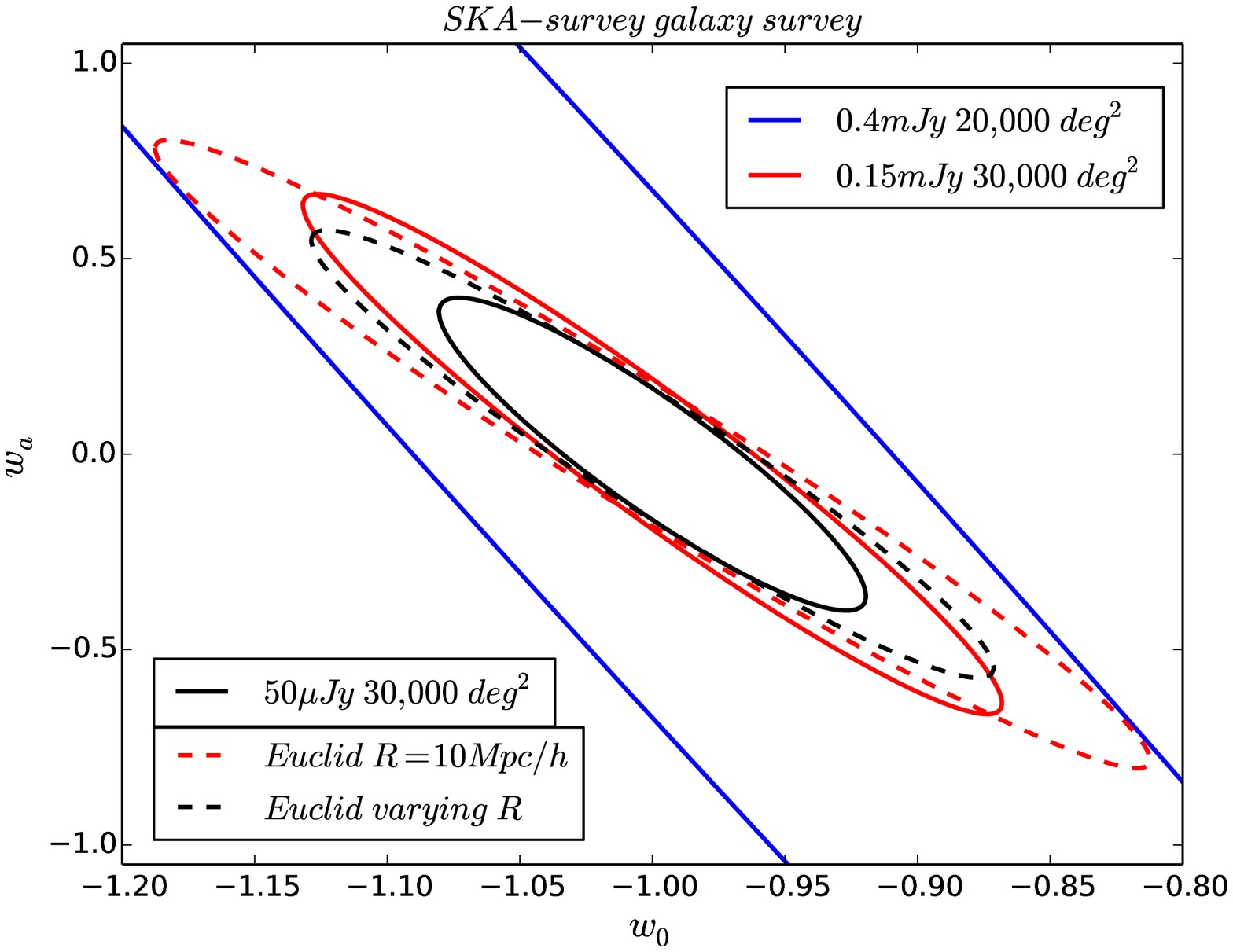}
\end{center}
\caption{\label{fig:DE_gal} Constraint on dark energy with HI galaxy survey. 
{\it Left}: HI galaxy survey with SKA1-mid, Right: HI galaxy survey with SKA1-survey.
The dashed contours indicate the constraints from Euclid. }
\end{figure}

\begin{figure}[!htbp]
\begin{center}
\includegraphics[width=0.49\textwidth]{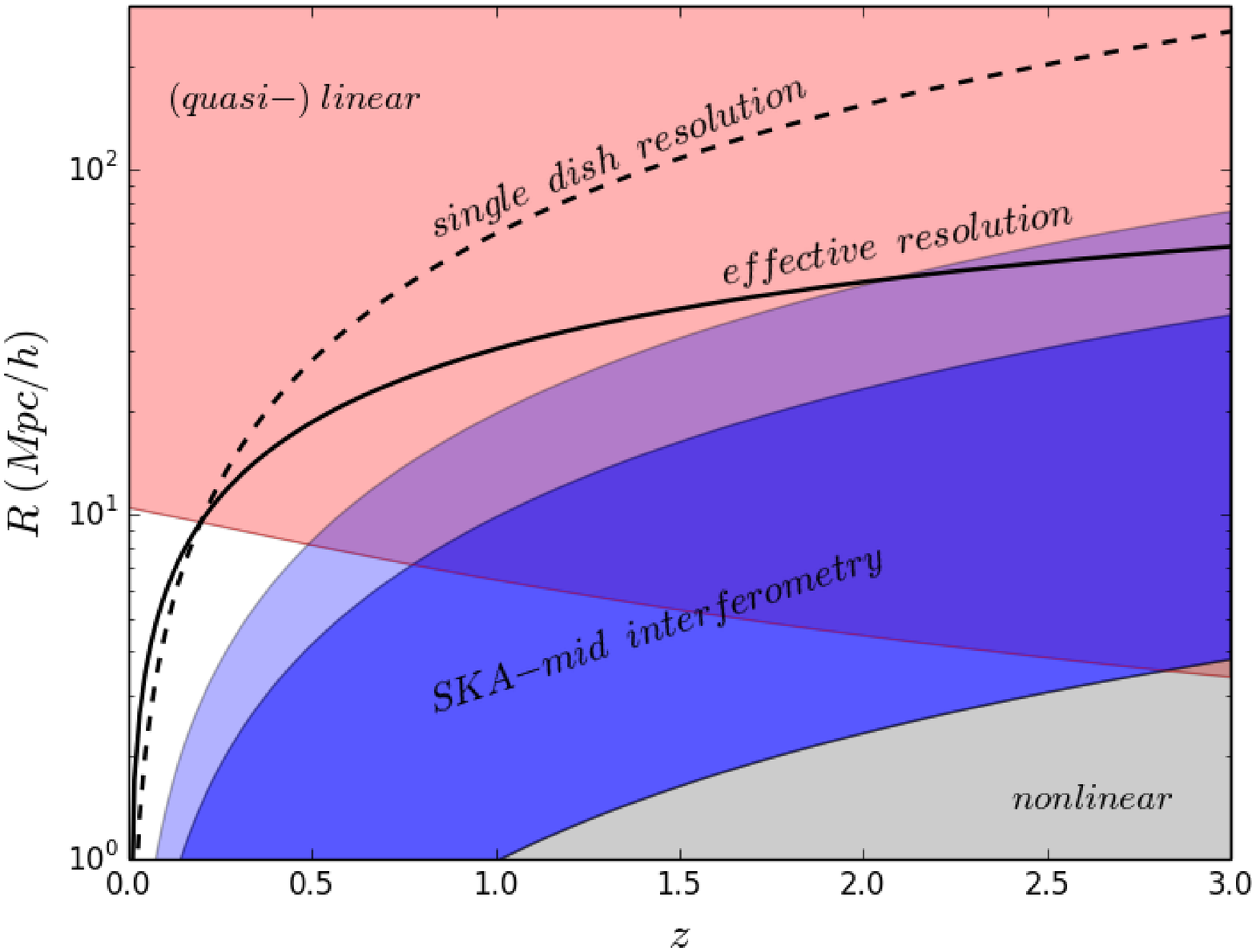}
\includegraphics[width=0.49\textwidth]{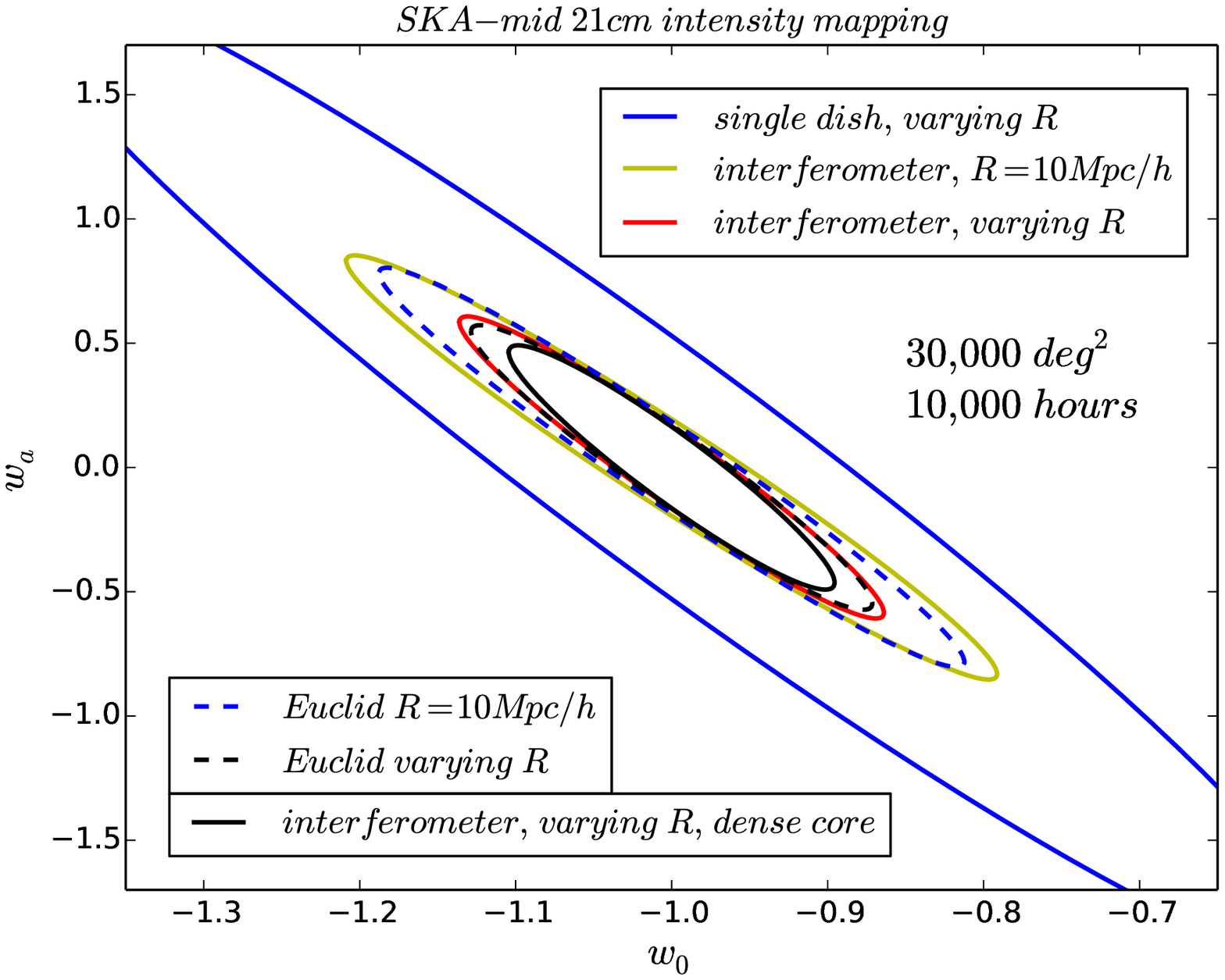}
\end{center}
\caption{\label{fig:SKA_mid_intensity} 
SKA-mid HI intensity mapping. {\it Left:} Various comoving scales as a function of redshift. 
The upper red region denotes the quasi-linear regime where the systematics of 
genus statistics are well-controlled.
Blue bands are most sensitive scales that would be observed by the core of SKA-mid
 interferometer, where we assume the baseline range from $100m$ ($50m$ for 
lighter blue band) to $1000m$. 
The grey region at bottom-right corner are scales available with the SKA spiral arm 
baselines. 
The black dashed line indicates the physical angular resolution of single dish. 
The solid black line is the effective resolution $(R_{ang}^2 R_{freq})^{1/3}$, where 
we utilize the minimum quasilinear scale, i.e. the boundary of red region, as $R_{freq}$.
{\it Right:} Constraints on dark energy equation of state $w_0$ and $w_a$. 
In the calculation, `varying R' means the smoothing length is selected 
using smallest quasi-linear scale. 
The dashed contours indicate the constraints from Euclid.}
\end{figure}

The HI galaxies survey constraints are plotted in 
Fig.\ (\ref{fig:DE_gal}), and IM survey constraints are plotted in the right panel of 
Fig.\ref{fig:SKA_mid_intensity}.
With a conservative assumption of the survey parameters that 
require $5,000$ hours integration time, neither constraint is very stringent for SKA1, 
but it is independent of other techniques, and strong constraints can be achieved 
with SKA2.  As we already know that BAO gives better constraints than genus for dark energy, we just give the 
constraints of our calculation, and refer the interested reader to the chapter written by Bull et al.. 

For the IM, there is certain complications. Unlike the BAO technique, 
where the physical scale is fixed, the topological standard ruler measure the scale 
around the smoothing length. To achieve better statistics, 
smaller smoothing length is desirable, though the systematics such as nonlinear growth put 
limits on the usable scale. In the left panel of Fig.\ref{fig:SKA_mid_intensity}), 
we illustrate various physical scales for SKA-mid. The quasi-linear 
region where systematics is well controlled is indicated by the upper red region.

For the single dish observing mode, the angular resolution by the $15m$ 
dish is poor (black dashed-line), varying
from $30\Mpc/h$ at $z=0.5$ to above $200\Mpc/h$ at $z=3$. However, 
in the line-of-sight direction the resolution is good, the effective $R_{\rm eff}$, 
\begin{eqnarray}
	R_{\rm eff} = (R^2_{\rm ang} R_{\rm freq})^{1/3}
\end{eqnarray}
becomes much smaller. Assuming a reasonable $R_{\rm freq}$ as indicated by 
the lower edge of the red region, the effective smoothing length 
 varies from $10\Mpc/h$ to $60\Mpc/h$. The cosmological constraints of 
this observation mode is shown as the blue contour in the right panel.

If sufficient number of short baselines is available, it may be possible to make
an IM survey in the interferometric mode. We assume a compact core in the center
of SKAmid, the blue belt in the left panel illustrates the most sensitive physical 
scale of this core, where we assume baselines distributed 
from 100m (50m for lighter blue region) to 1000m. 
With the systematic limit from nonlinear scale, one would be 
able to observe higher redshift volume from $z=0.78$ ($z=0.5$) at much 
smaller smoothing length and therefore a better 
statistical accuracy. As shown in the right panel, such observation 
mode improve the constraints significantly.

For modified gravity models, we consider specifically the $f(R)$ theory with a single
model parameter $B_0$ characterizing the present value of Compton scale.
HI galaxy survey by SKA2 will be able to constrain $B_0$ to $5\times 10^{-5}$ 
at 1-$\sigma$ level for a fiducial value of $B_0=10^{-4}$, while 
the best constraints from intensity mapping is $8\times 10^{-5}$. This is much better than the current strongest constraints of 
$1.1\times 10^{-3}$ at $95\%$ C.L..

\section{Reionization}

\begin{figure}
\begin{center}
\includegraphics[width=0.75\textwidth]{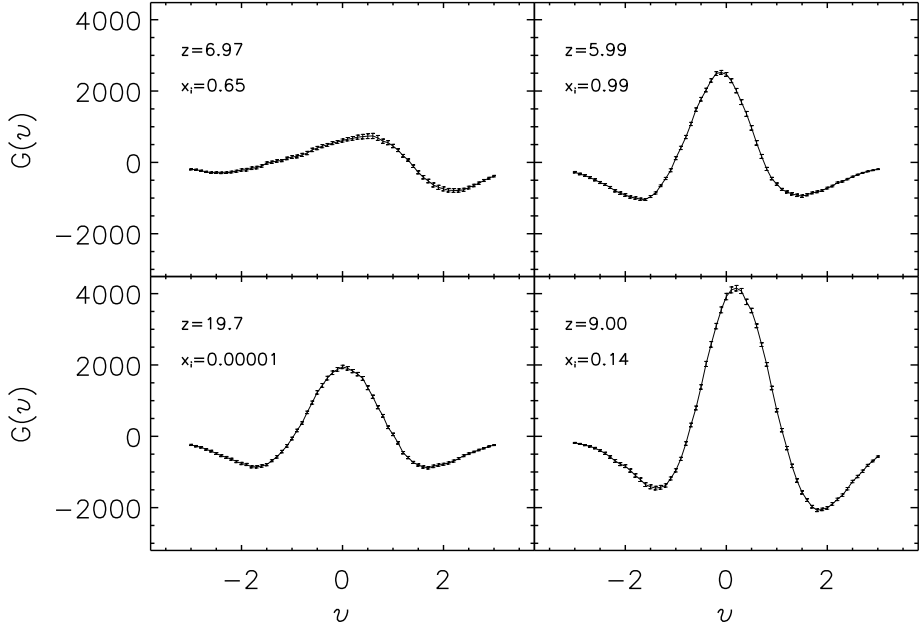}
\end{center}
\caption{Genus curve and statistical error of $\delta T_b$ at different redshifts from a 
simulation (Trac et al. 2008), 
the simulation box has a side length of $100 {\rm Mpc\ h^{-1}}$, and smoothing 
length is $1 {\rm Mpc\ h^{-1}}$
}
\label{fig:3dgenus_error}
\end{figure}

The genus curve can be used to distinguish the different phases of 
reionization \citep{2008ApJ...675....8L,2014JKAS...47...49H} discussed in \S \ref{intro:reion}. 
In Fig.~\ref{fig:3dgenus_error} we show the genus curve from an EoR 
simulation \citep{2008ApJ...689L..81T}, for four redshifts during the EoR.
As expected, at the highest redshift($z=19.7, x_i\approx 0$), the genus curve 
has the shape for that of a Gaussian distribution. The amplitude of the curve increases
in the pre-overlap phase ($z=9,x_i=0.14$), but retains the shape for Gaussian distribution.
The shape of the genus curve is drastically different during the overlap phase ($z=6.97,
x_i=0.65$). Finally, the curve returns to the shape of Gaussian distribution in the 
post-reionization phase ($z=5.99,x_i=0.99$). 
From Eq.\ref{eqn:genus_A} we see the amplitude of genus curve ${A} \propto k_e^3$,
$k_e$ first increases as the HII bubbles grows in number, and then
decreases as the bubbles merge and overlap. Measuring $k_e$ from ${A}$ can provide
information on the bubble sizes.

These curves are obtained with a smoothing length of $1\Mpc/h$. Along with the genus curves, 
the statstical error obtained from a $100 \Mpc/h$ simulation box is plotted.
The design of the SKA-low (Dewdney et al. 2013; Braun 2013) is still very uncertain, 
but it is agreed that there should be a compact core within 1km which provides 
high sensitivity observation
on scales of a few arcmin for redshift 10. This is roughly sufficient for the 
measurement. The observation of SKA-low for a deep field should be able to 
yield data which can be compared with such simulations.

\section{Conclusion}
Topology can be very useful at characterizing random fields. The SKA, with its high sensitivity, can observe the HI distribution much beyond 
the current limits. We can apply the topological analysis method to the SKA HI survey
data. Such analysis can also be used to test or constrain primordial non-Gaussianity,
dark energy and modified gravity.  The constraints for the model parameters from the topological method are not as tight as the traditional 
methods, e.g., dark energy constraints from BAO measurements (See Bull et al. chapter) and primordial non-Gaussianity from scale-dependent bias and 
bispectrum measurements (See Camera et al. chapter), but it is very robust 
against non-linearity, bias and redshift distortion in the evolution of 
large scale structure.  
During the EoR, it can also distinguish different evolution phases and characters of
reionization model. We made forecasts for some of these applications, in both 
HI galaxy survey and HI intensity mapping survey, and obtained constraints for different
survey parameters. For SKA1, the precision of the measurement with the genus curve 
is moderate, but we anticipate that the addition of other Minkowski functionals 
would help improve these constraints, and for SKA2 the result is competitive. More 
importantly, the genus curve is capable of probing non-Gaussian features which are not
parameterized in the standard form, so with SKA it would be 
an important tool for probing and discovering non-Gaussianity in unexplored regimes.

\bibliographystyle{apj}
\bibliography{ms}

\begin{thebibliography}{22}
\expandafter\ifx\csname natexlab\endcsname\relax\def\natexlab#1{#1}\fi

\bibitem[{{Abdalla} \& {Rawlings}(2005)}]{2005MNRAS.360...27A}
{Abdalla}, F.~B. \& {Rawlings}, S. 2005, \mnras, 360, 27

\bibitem[{{Blake} {et~al.}(2014){Blake}, {James}, \&
  {Poole}}]{2014MNRAS.437.2488B}
{Blake}, C., {James}, J.~B., \& {Poole}, G.~B. 2014, \mnras, 437, 2488

\bibitem[{{Braun}(2013)}]{} 
{Braun}, R. 2013,``\rm{SKA1 Imaging Science Performance}",
SKA-TEL-SKO-DD-XXX, Revision A Draft 2


\bibitem[{{Chang} {et~al.}(2008){Chang}, {Pen}, {Peterson}, \&
  {McDonald}}]{2008PhRvL.100i1303C}
{Chang}, T.-C., {Pen}, U.-L., {Peterson}, J.~B., \& {McDonald}, P. 2008,
  Physical Review Letters, 100, 091303

\bibitem[{{Choi} {et~al.}(2013){Choi}, {Kim}, {Rossi}, {Kim}, \&
  {Lee}}]{2013ApJS..209...19C}
{Choi}, Y.-Y., {Kim}, J., {Rossi}, G., {Kim}, S.~S., \& {Lee}, J.-E. 2013,
  \apjs, 209, 19

\bibitem[{{Choi} {et~al.}(2010){Choi}, {Park}, {Kim}, {Gott}, {Weinberg},
  {Vogeley}, {Kim}, \& {SDSS Collaboration}}]{2010ApJS..190..181C}
{Choi}, Y.-Y., {Park}, C., {Kim}, J., {et~al.} 2010, \apjs, 190, 181

\bibitem[{{Dewdney}{et~al.}(2013){Dewdney},{Turner},{Millenaar},{McCool},{Lazio}, \&{Cornwell}}]{}
{Dewdney}, P.~E., {Turner}, W.,  {Millenaar}, R.,
 {McCool}, R.,  {Lazio}, J., \&  {Cornwell}, T.~J. 2013, ``\rm{SKA1 System Baseline Design}",
\rm{SKA-TEL-SKO-DD-001}, Revision 1 


\bibitem[{{Gott} {et~al.}(1986){Gott}, {Dickinson}, \&
  {Melott}}]{1986ApJ...306..341G}
{Gott}, III, J.~R., {Dickinson}, M., \& {Melott}, A.~L. 1986, \apj, 306, 341

\bibitem[{{Hamilton} {et~al.}(1986){Hamilton}, {Gott}, \&
  {Weinberg}}]{1986ApJ...309....1H}
{Hamilton}, A.~J.~S., {Gott}, III, J.~R., \& {Weinberg}, D. 1986, \apj, 309, 1

\bibitem[{{Hikage} {et~al.}(2008){Hikage}, {Coles}, {Grossi}, {Moscardini},
  {Dolag}, {Branchini}, \& {Matarrese}}]{2008MNRAS.385.1613H}
{Hikage}, C., {Coles}, P., {Grossi}, M., {et~al.} 2008, \mnras, 385, 1613

\bibitem[{{Hikage} {et~al.}(2006){Hikage}, {Komatsu}, \&
  {Matsubara}}]{2006ApJ...653...11H}
{Hikage}, C., {Komatsu}, E., \& {Matsubara}, T. 2006, \apj, 653, 11

\bibitem[{{Hong} {et~al.}(2014){Hong}, {Ahn}, {Park}, {Kim}, {Iliev}, \&
  {Mellema}}]{2014JKAS...47...49H}
{Hong}, S.~E., {Ahn}, K., {Park}, C., {et~al.} 2014, Journal of Korean
  Astronomical Society, 47, 49

\bibitem[{{Lee} {et~al.}(2008){Lee}, {Cen}, {Gott}, \&
  {Trac}}]{2008ApJ...675....8L}
{Lee}, K.-G., {Cen}, R., {Gott}, III, J.~R., \& {Trac}, H. 2008, \apj, 675, 8

\bibitem[{{Matsubara}(2003)}]{2003ApJ...584....1M}
{Matsubara}, T. 2003, \apj, 584, 1

\bibitem[{{Mecke} {et~al.}(1994){Mecke}, {Buchert}, \&
  {Wagner}}]{1994A&A...288..697M}
{Mecke}, K.~R., {Buchert}, T., \& {Wagner}, H. 1994, \aap, 288, 697

\bibitem[{{Myers} {et~al.}(2009){Myers}, {Abdalla}, {Blake}, {Koopmans},
  {Lazio}, \& {Rawling}}]{2009astro2010S.219M}
{Myers}, S.~T., {Abdalla}, F.~B., {Blake}, C., {et~al.} 2009, in ArXiv
  Astrophysics e-prints, Vol. 2010, astro2010: The Astronomy and Astrophysics
  Decadal Survey, 219

\bibitem[{{Park} {et~al.}(2005){Park}, {Kim}, \& {Gott}}]{2005ApJ...633....1P}
{Park}, C., {Kim}, J., \& {Gott}, III, J.~R. 2005, \apj, 633, 1

\bibitem[{{Park} \& {Kim}(2010)}]{2010ApJ...715L.185P}
{Park}, C. \& {Kim}, Y.-R. 2010, \apjl, 715, L185

\bibitem[{{Park} {et~al.}(2013){Park}, {Pranav}, {Chingangbam}, {van de
  Weygaert}, {Jones}, {Vegter}, {Kim}, {Hidding}, \&
  {Hellwing}}]{2013JKAS...46..125P}
{Park}, C., {Pranav}, P., {Chingangbam}, P., {et~al.} 2013, Journal of Korean
  Astronomical Society, 46, 125

\bibitem[{{Pratten} \& {Munshi}(2012)}]{2012MNRAS.423.3209P}
{Pratten}, G. \& {Munshi}, D. 2012, \mnras, 423, 3209

\bibitem[{{Santos}(2014)}]{}
Santos, M., \emph{Cosmology with the SKA}, talk at
Cosmology Science Workgroup in Jan 2014



\bibitem[{{Speare} {et~al.}(2013){Speare}, {Gott}, {Kim}, \&
  {Park}}]{2013arXiv1310.4278S}
{Speare}, R., {Gott}, J.~R., {Kim}, J., \& {Park}, C. 2013, ArXiv e-prints

\bibitem[{{Trac} {et~al.}(2008){Trac}, {Cen}, \& {Loeb}}]{2008ApJ...689L..81T}
{Trac}, H., {Cen}, R., \& {Loeb}, A. 2008, \apjl, 689, L81

\bibitem[{{Wang} {et~al.}(2012){Wang}, {Chen}, \& {Park}}]{2012ApJ...747...48W}
{Wang}, X., {Chen}, X., \& {Park}, C. 2012, \apj, 747, 48

\bibitem[{{Zunckel} {et~al.}(2011){Zunckel}, {Gott}, \&
  {Lunnan}}]{2011MNRAS.412.1401Z}
{Zunckel}, C., {Gott}, J.~R., \& {Lunnan}, R. 2011, \mnras, 412, 1401

\end{thebibliography}

\end{document}